\documentclass{iau}

\usepackage{amsmath}
\usepackage{graphicx}
\usepackage{multirow}
\usepackage[export]{adjustbox}
\usepackage{caption}
\usepackage{subcaption}

\begin{document}

\jnlPage{1}{7}
\jnlDoiYr{2021}
\doival{10.1017/xxxxx}

\aopheadtitle{Proceedings IAU Symposium}
\editors{C. Sterken,  J. Hearnshaw \&  D. Valls-Gabaud, eds.}

\title{Exploring connections between the VLBI and optical morphology of AGNs and their host galaxies}

\author{David Fernández Gil, Jeffrey A. Hodgson, Benjamin L'Huillier}
\affiliation{Sejong University, Dept. of Physics and Astronomy, Seoul, South Korea}

\begin{abstract}
We analyse VLBI and optical images of AGNs and their host galaxies and look for statistical correlations between the shape and orientation of the galaxy and the direction of the jet. We utilise the Astrogeo catalogue, which has over 9000 VLBI sources, many of those with a clear core-jet like structure that allows for the jet position angle to be reliably determined. We then use the VLBI source positions to search for optical counterparts within various optical surveys. In order to parameterise the orientation and shape of the host galaxy, we fitted a Gaussian elliptical model to the optical image, taking the PSF into account. We check our own shape parameters from this fit against the ones provided by the optical surveys. As of yet, no clear correlation between the galaxy morphology and the jet direction is seen. 
\end{abstract}


\maketitle

\section{Introduction}

Active Galactic Nuclei (AGN) are compact central regions of galaxies that can produce relativistic jets. These jets, when pointed towards Earth, are referred to as blazars and can be detected via VLBI. AGN play an active role in star-formation feedback and are useful for studying distant galaxies due to their high brightness (\cite{1995PASP..107..803U}). \\

The study aims to investigate the correlation between small-scale jet properties and large-scale host galaxy characteristics, specifically examining the correlation between jet orientation and host galaxy optical shape. The expectation is that the jets may be perpendicular to the plane of the host galaxy; however, factors such as interaction with the environment, energy changes, projection and relativistic effects can complicate this relationship. The goal is to use a large sample to determine if there are any statistically significant generalities in the orientation of the jets relative to their host galaxies.

\section{Observations}

We utilize the Astrogeo VLBI catalogue, a geodetic catalogue, to gather data from over 9000 VLBI sources. The catalogue includes the coordinates of the sources and the jet position angle (PA) (\cite{2022ApJS..260....4P}). The optical counterpart and shape parameters of the host galaxies are obtained from several optical surveys, indicated in Table \ref{Tab1}.

\begin{table}[h!]
\centering
\caption{Information about the optical surveys and their shape parameters.}\label{Tab1}
{\tablefont\begin{tabular}{@{\extracolsep{\fill}}lllllll}
\midrule
Survey Name & Median Seeing & Band & Good fit definition & Reference\\
\midrule
Sloan Digital Sky Survey (SDSS) & 1.43'' & r & type `GALAXY' & \cite{2022ApJS..259...35A} \\
Dark Energy Spectroscopic Instrument (DESI) & 1.2" & g & type `Sérsic' & \cite{2019AJ....157..168D}\\
Kilo Degree Survey Golden Catalogue (KiDS) & 0.7-0.9" & g & b $>$ 2px, stellar index $<$ 0.5 & \cite{2013Msngr.154...44D} \\
Skymapper & $~$2.6" & g & b $>$ 2px, stellar index $<$ 0.5 & \cite{2009ASPC..404..356M}\\
\midrule
\end{tabular}}
\end{table}

\section{Methods}
 We perform a cross-match by using the VLBI coordinates as a reference and searching for the closest optical source within 5 arcseconds. Shape parameters are obtained from published optical surveys (refer to Table \ref{Tab1}). Aditionally, we download the images and PSF of the optical counterparts in SDSS and forward-model their brightness distribution performing our own simple elliptical Gaussian fit (\cite{2023arXiv230107725F}). This provides the PA of the major and minor axis of the galaxy and the axis ratio. The VLBI jet's PA is then compared to the minor axis PA of the fitted ellipse, as this axis represents the perpendicular direction to the projected galaxy image.

 \section{Results}
\begin{figure}[h]
  \centering
  \subfloat[Axis ratio.]{\includegraphics[width=0.45\textwidth]{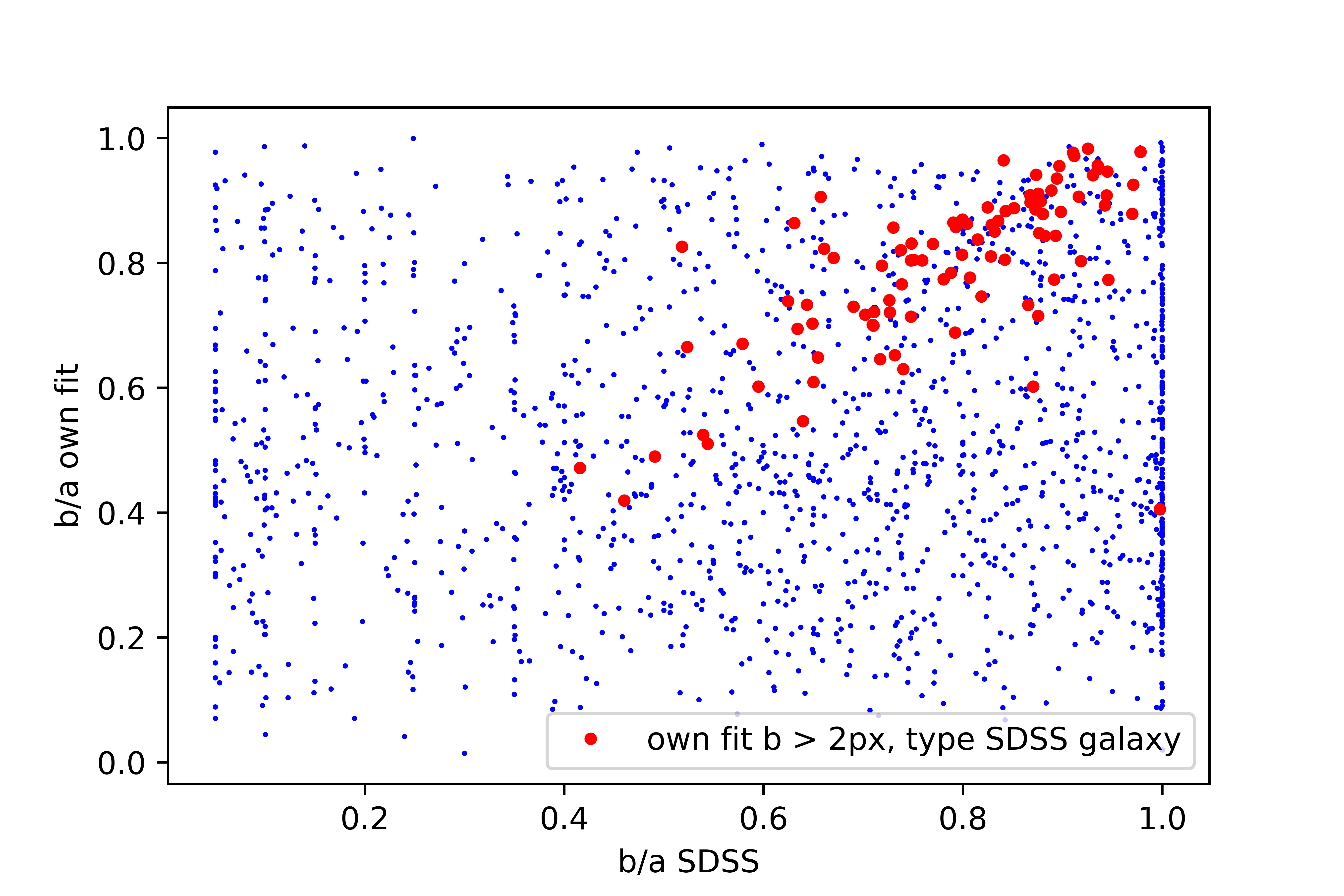}\label{Fig1.1}}
  \subfloat[Semi-major axis position angle.]{\includegraphics[width=0.45\textwidth]{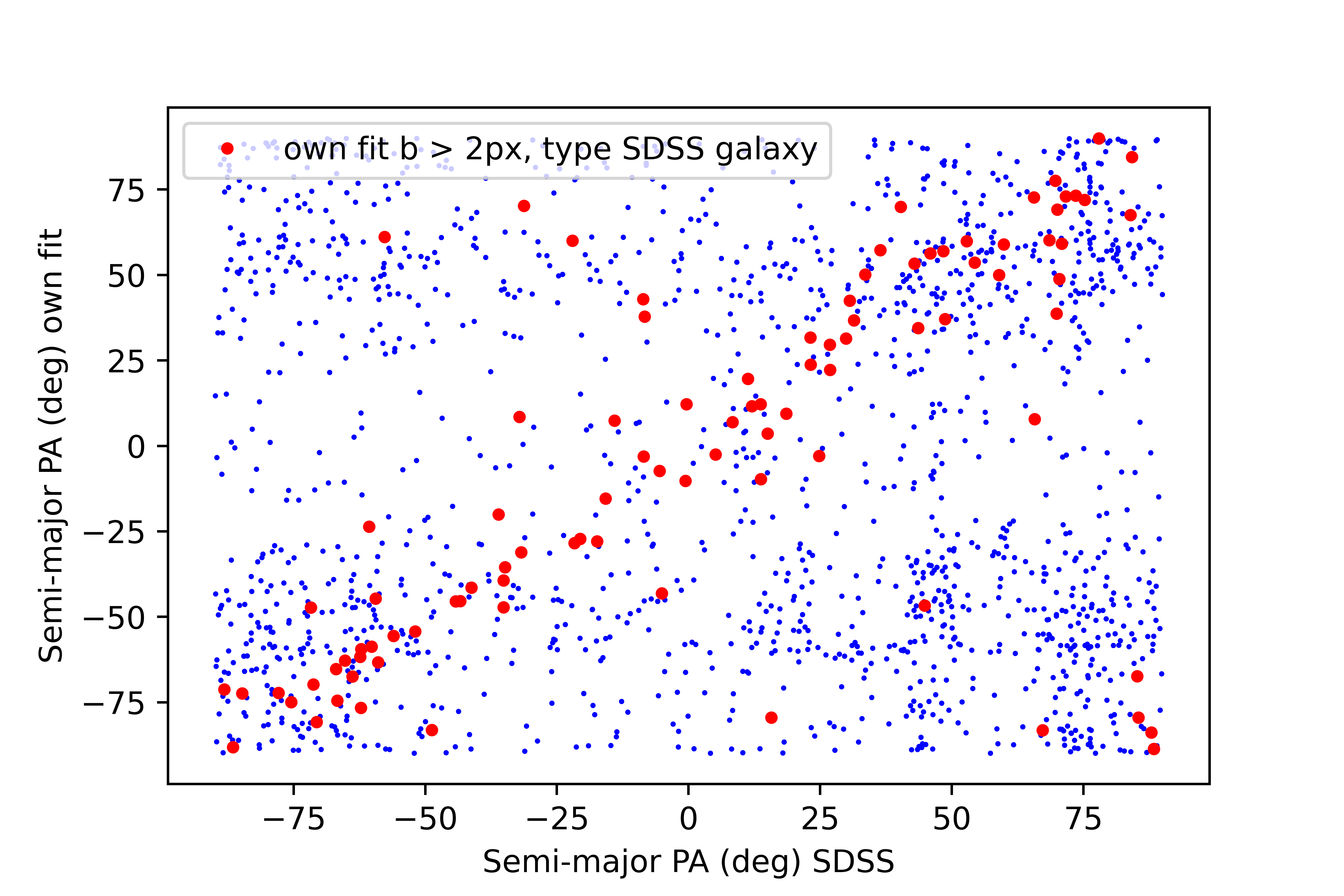}\label{Fig1.2}}
  \caption{Comparison between our Gaussian fits and SDSS shape parameters. In red are the cases where the semi-minor axis of our fit is greater than 2 pixels and SDSS classifies the fit as `Galaxy' type (resolved).}
\end{figure}

 We first compare the results of our Gaussian fit with the parameters provided by the SDSS collaboration (\cite{2022ApJS..259...35A}). We define a good fit as when the semi-minor axis is greater than 2 pixels. A good fit from the SDSS is defined as when the catalogue classifies the source as type `GALAXY' (resolved). In Figs. \ref{Fig1.1} and \ref{Fig1.2} we plot all fits (blue) and good fits (red) from both. We can see that the good fits are correlated. \\
 
\begin{figure}[h]
  \centering
  \subfloat[Axis ratio.]{\includegraphics[width=0.45\textwidth]{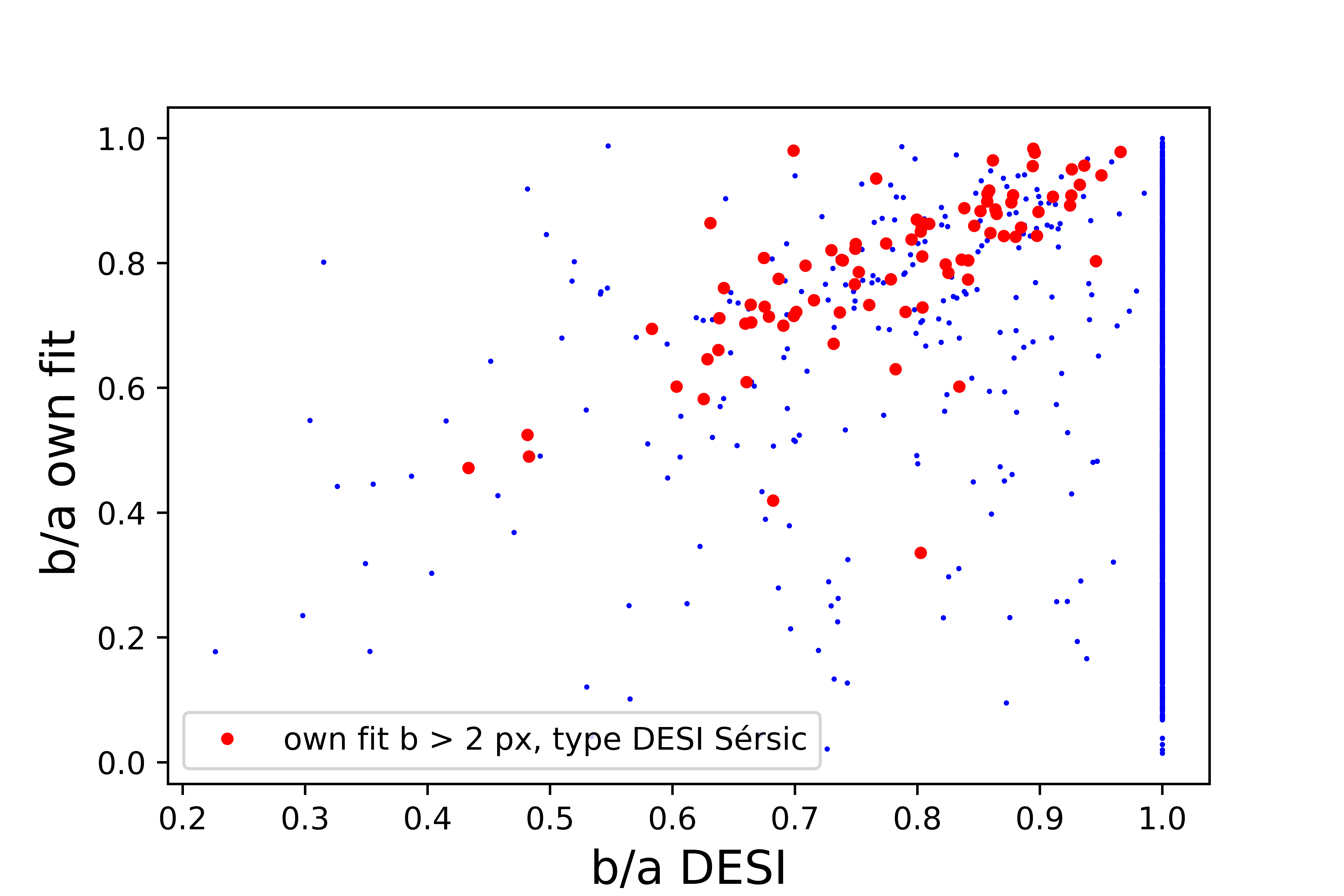}\label{Fig2.1}}
  \subfloat[Semi-major axis position angle.]{\includegraphics[width=0.45\textwidth]{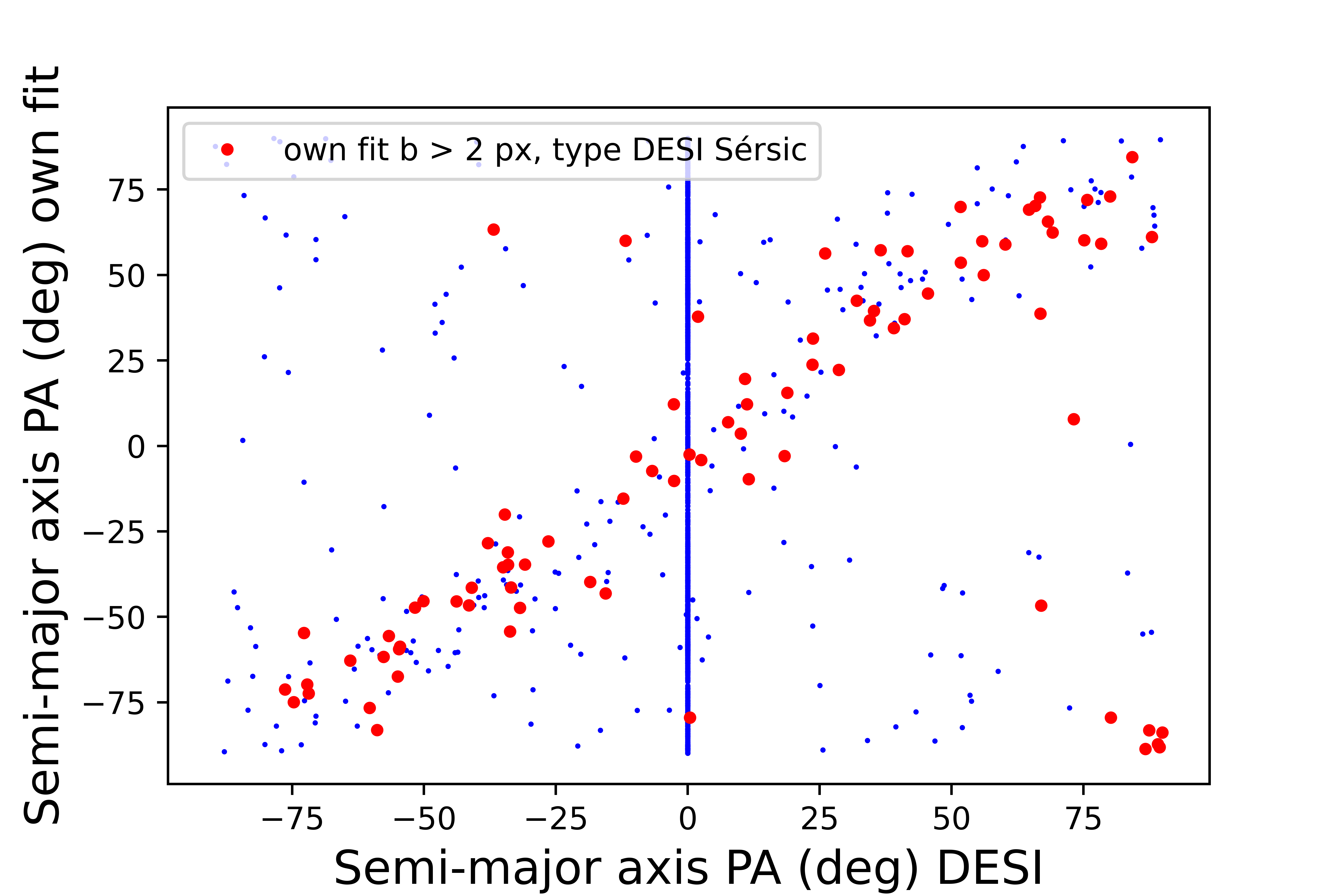}\label{Fig2.2}}
  \caption{Comparison between our Gaussian fits and DESI shape parameters. In red are the cases where the semi-minor axis of our fit is greater than 2 pixels and DESI classifies the fit as `Sérsic' type (resolved). The horizontal lines of blue dots just show many cases where DESI could not resolve the shape, so it automatically assigned a PA of zero and axis ratio of one.}
\end{figure}

 We then include optical shape data from the other optical surveys that are listed in Table \ref{Tab1}. For sources that are in both SDSS and DESI, we also compare our Gaussian fits with their shape parameters. A good fit in DESI is defined as when the shape model of the source is classified as 
`Sérsic'. As can be seen in Figs. \ref{Fig2.1} and \ref{Fig2.2}, the good fits are also correlated. The horizontal lines of blue dots just show many cases where DESI could not resolve the shape, so it just assigned a PA of zero and axis ratio of one.  \\ 
 
 We perform a similar analysis using KiDS and Skymapper and find that for the good fit cases the results are correlated (see Table \ref{Tab1} for the definition of a good fit in each survey).
 
 We find that the simple Gaussian fit performs well. However, there are some outliers. We individually inspect some of these cases and find that the outliers in the corners of Figure \ref{Fig1.2} correspond with cases where the 2-spin symmetry made the position angle of the semi-major axis jump from +90 to -90 even if the difference is slight. The other outliers are QSO objects which, although marked as resolved by the catalogues, are actually not well resolved and could not be fit correctly.

\begin{figure}[h]
  \centering
  \subfloat[SDSS.]{\includegraphics[width=0.45\textwidth]{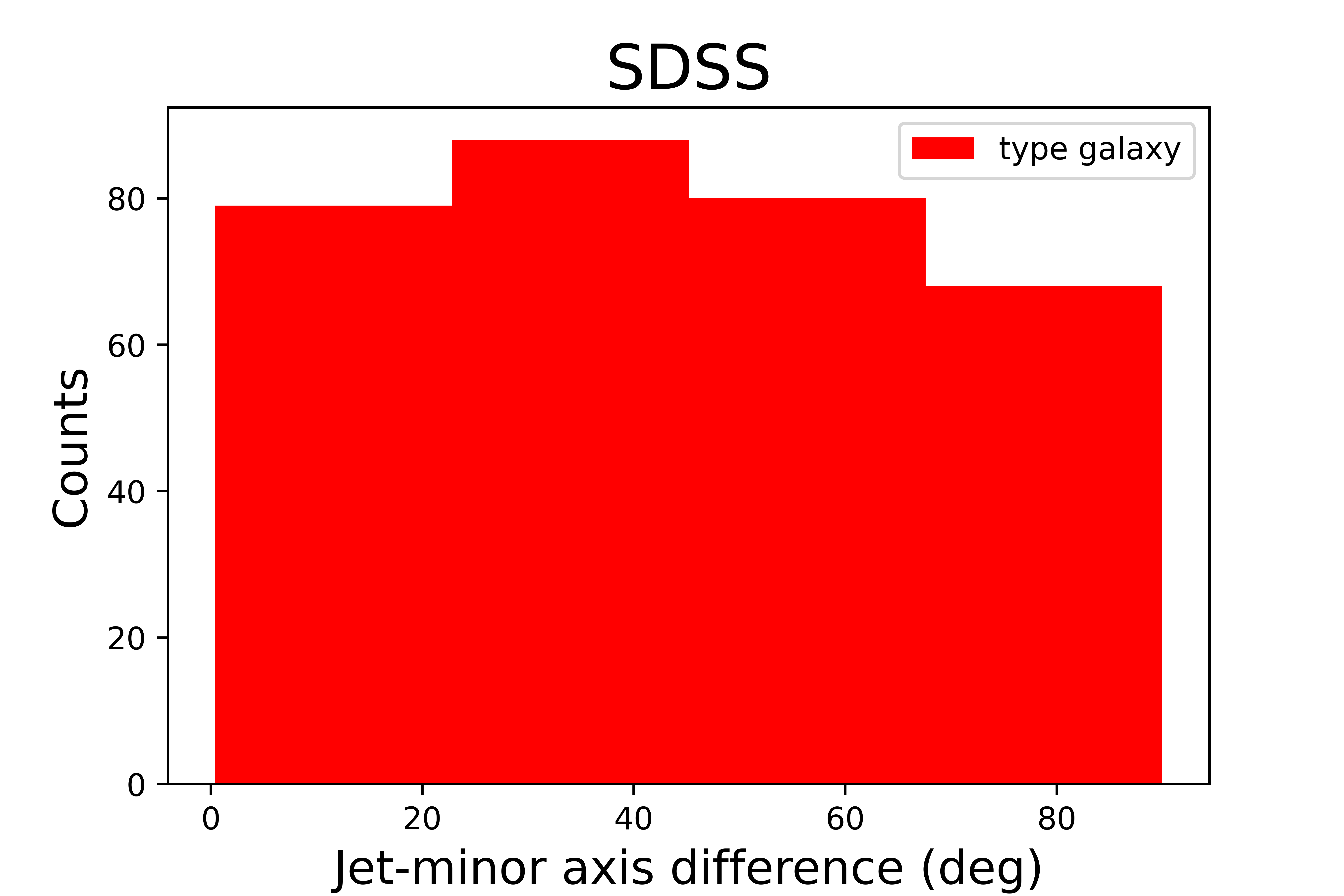}\label{Fig3.1}}
  \subfloat[DESI.]{\includegraphics[width=0.45\textwidth]{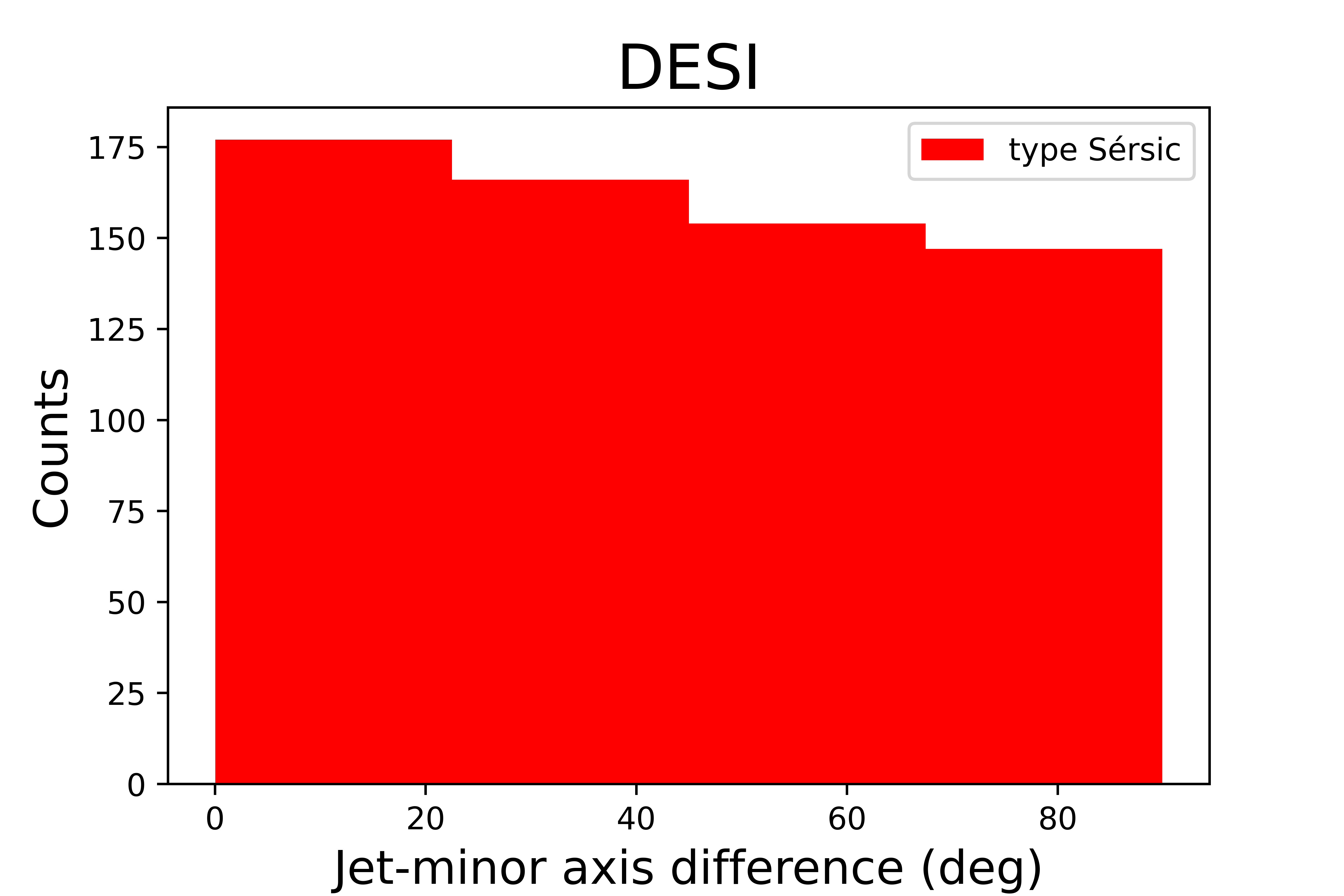}\label{Fig3.2}} \\
  \subfloat[Skymapper.]{\includegraphics[width=0.45\textwidth]{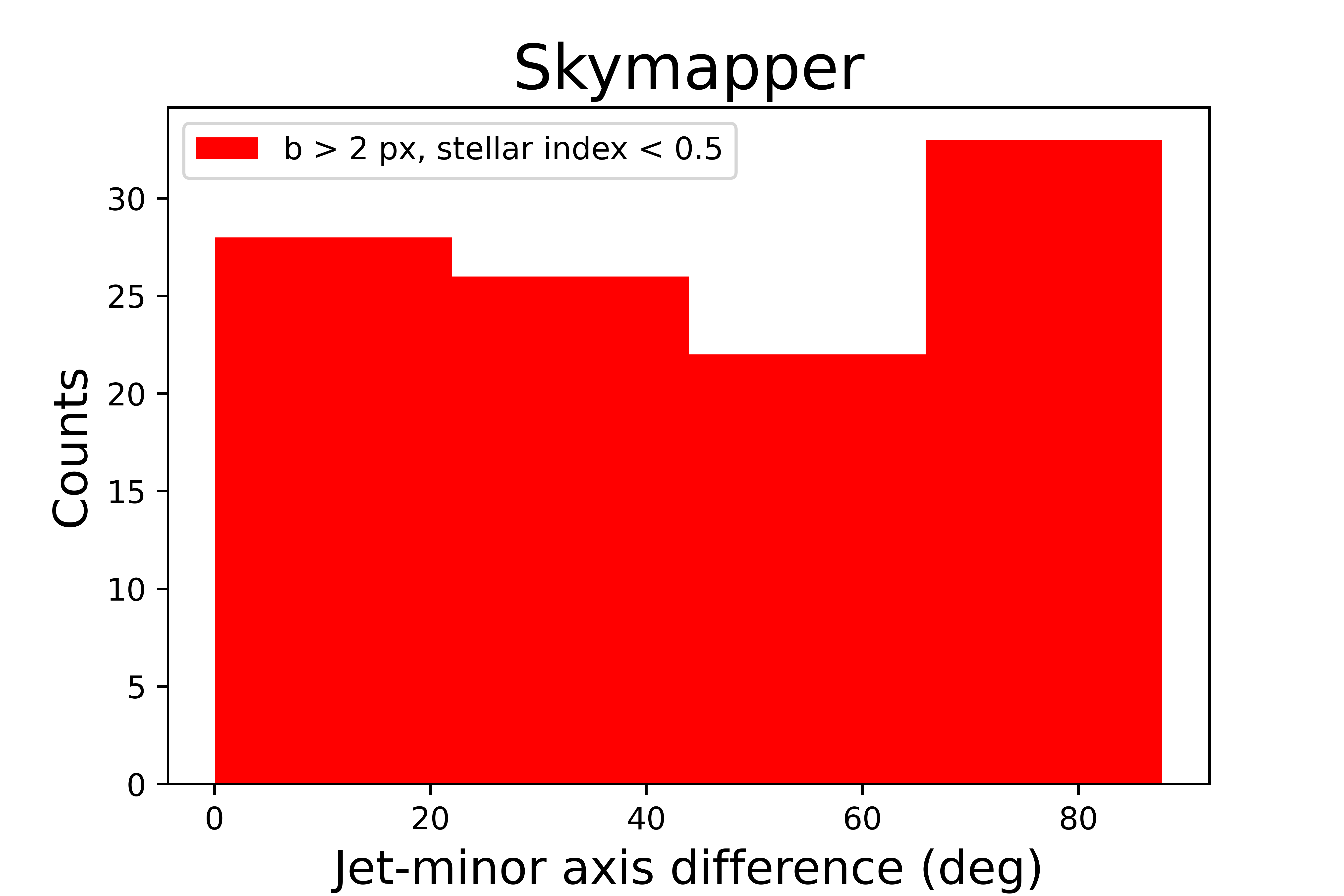}\label{Fig3.3}}
  \subfloat[KiDS.]{\includegraphics[width=0.45\textwidth]{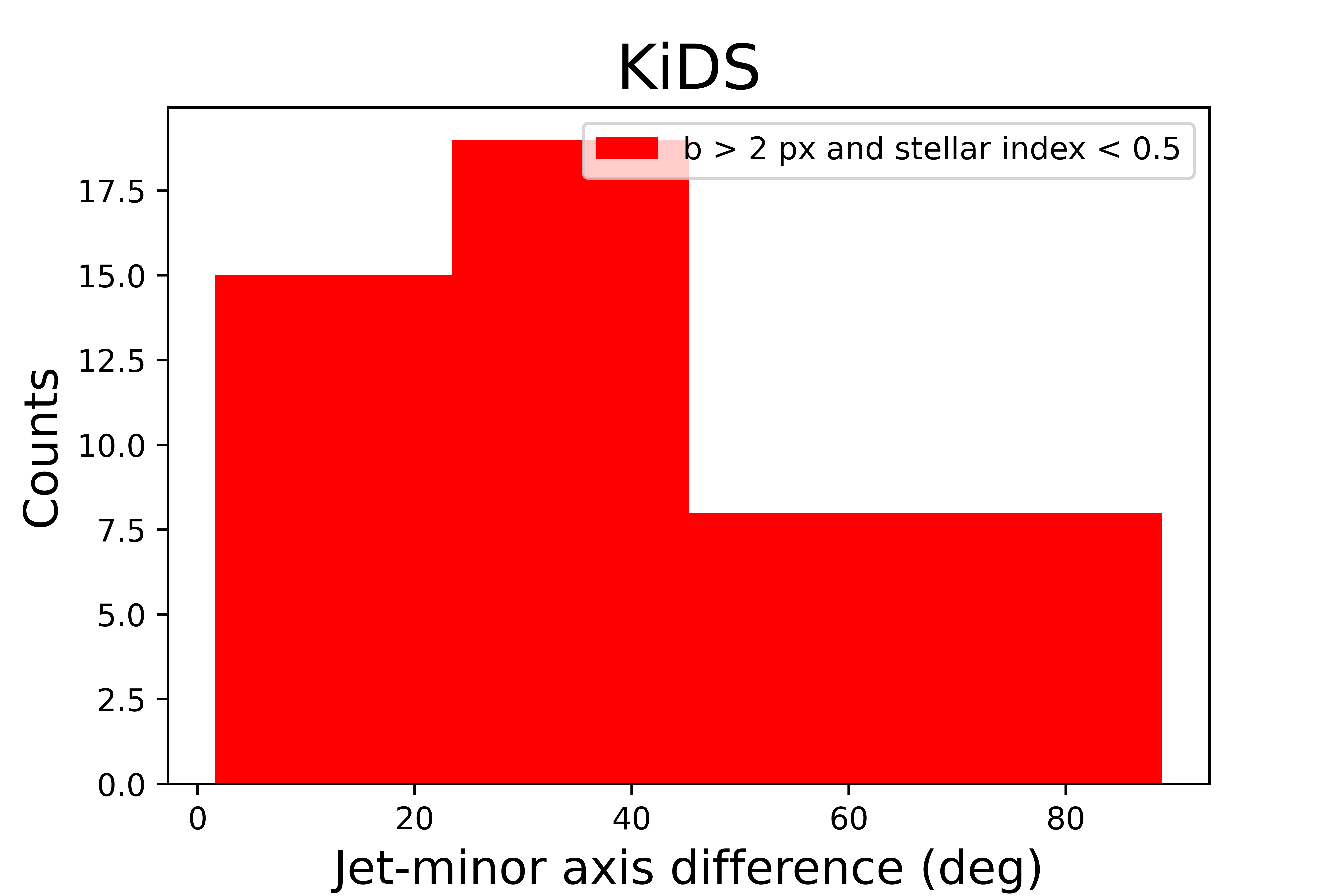}\label{Fig3.4}}
  \caption{Angle between the jet of the AGN and the semi-minor axis of the fitted shape of the host galaxies by the different optical catalogues, only including the cases where the fit was good, according to our criteria in Table \ref{Tab1}.}
\end{figure}

\begin{figure}[h]
    \includegraphics[width=0.5\textwidth, center]{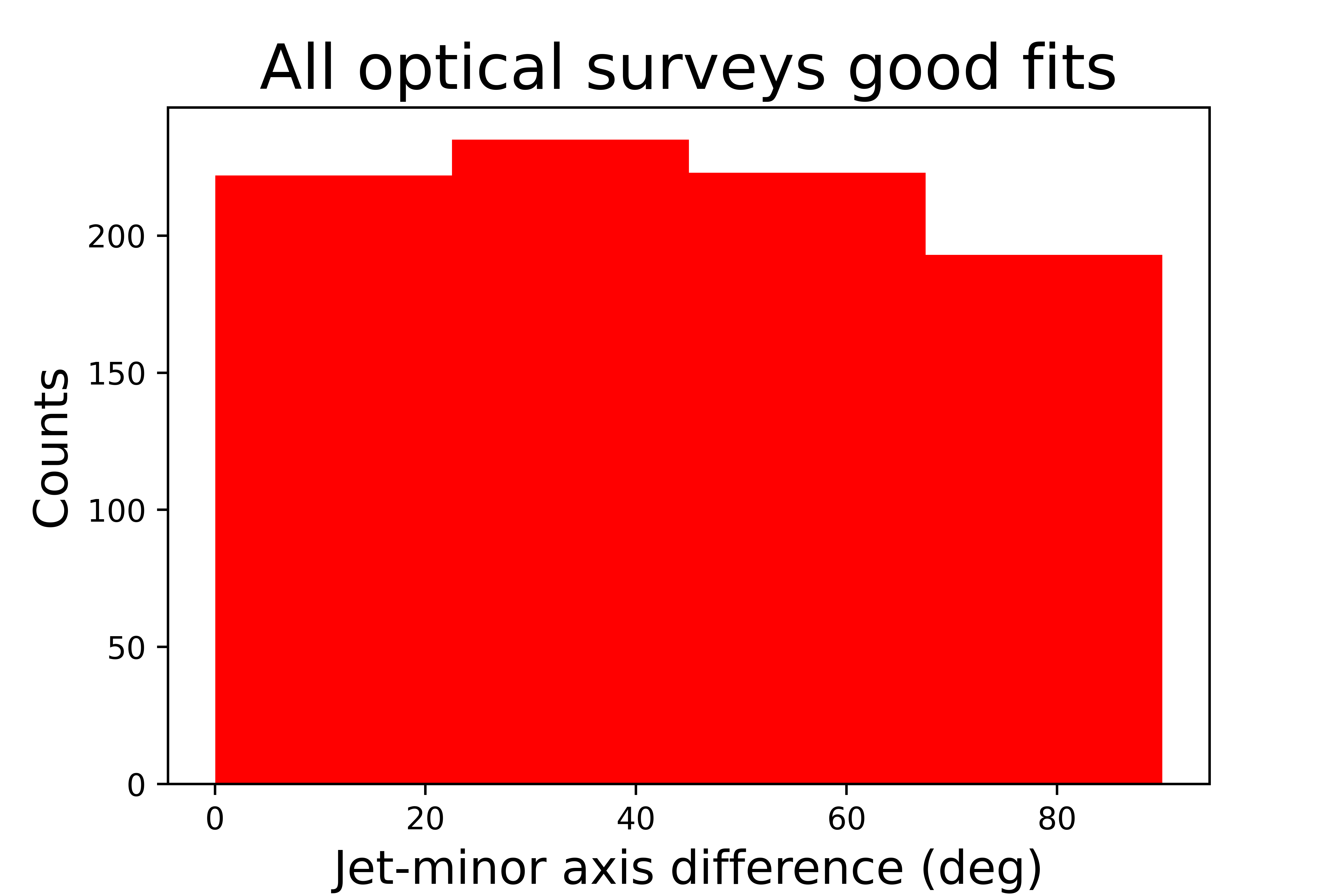}
    \caption{Combined histogram from the ones in Figs. \ref{Fig3.1}, \ref{Fig3.2}, \ref{Fig3.3} and \ref{Fig3.4}.}
    \label{Fig4}
\end{figure}

Figs. \ref{Fig3.1}, \ref{Fig3.2}, \ref{Fig3.3} and \ref{Fig3.4} contain histograms with the angle between the jet and the semi-minor axis of the galaxies for each database, only using the 'good fit' cases as defined previously and in Table \ref{Tab1}. In Figure \ref{Fig4}, we plot a combined histogram with all surveys together, using the mean value of the angle difference if the source is present in several surveys.

\section{Discussion and conclusions}
Overall, no clear angle preference for the jet is found. However, it is interesting that there may be hints of a signal in the highest resolution data (e.g. DESI and KiDS). Nevertheless at this stage no strong claims can be made. Adding more high resolution optical data (e.g. Dark Energy Survey and/or Hubble Space Telescope) may ultimately resolve the question. In any case, even if a signal is not detected, this still has important implications for fields as diverse as stellar and extra-galactic astrophysics.

\end{document}